\documentclass[journal]{IEEEtran} 

\usepackage{cite}

\ifCLASSINFOpdf
\else
\fi
\ifCLASSOPTIONcompsoc
  \usepackage[caption=false,font=normalsize,labelfont=sf,textfont=sf]{subfig}
\else
  \usepackage[caption=false,font=footnotesize]{subfig}
\fi
\usepackage{url}

\usepackage{float}
\usepackage{amssymb}
\usepackage{amsmath}
\usepackage{amsthm}
\usepackage{tikz}
\usepackage{graphicx}
\usepackage{pgfplots}
\usepackage{pstool}  
\usepackage[linesnumbered,algoruled,boxed,lined]{algorithm2e}
\usepackage{algcompatible}
\usepackage{threeparttable}
\usepackage[strict]{changepage}
\usepackage[keeplastbox]{flushend}
\usepackage{bbm}
\usepackage{bm}
\usepackage{comment}
\usepackage{hhline}
\usepackage{enumitem}

\pgfplotsset{compat=1.14}

\begin{document}

%
\title{Deep Learning in Downlink Coordinated Multipoint in New Radio Heterogeneous Networks}


\author{Faris B.~Mismar,~\IEEEmembership{Senior Member,~IEEE}
        and~Brian L.~Evans,~\IEEEmembership{Fellow,~IEEE}
\thanks{The authors are with the Wireless Networking and Communications Group, Dept. of Electrical and Comp. Eng., The University of Texas at Austin, Austin, TX, 78712, USA e-mail: faris.mismar@utexas.edu,  bevans@ece.utexas.edu.} %
}

\newcommand{\dif}{\mbox{d}} 
\newcommand{\argmax}{\operatorname{arg\, max\,}}
\newcommand{\argmin}{\operatorname{arg\, min\,}}
\newcommand{\given}{\,\vert\,}
\newcommand{\Given}{\,\bigg\vert\,}

\maketitle

\begin{abstract}
We propose a method to improve the performance of the downlink coordinated multipoint (DL CoMP) in heterogeneous fifth generation New Radio (NR) networks. The standards-compliant method is based on the construction of a surrogate CoMP trigger function using deep learning.  The cooperating set is a single-tier of sub-6 GHz heterogeneous base stations operating in the frequency division duplex mode (i.e., no channel reciprocity).  This surrogate function enhances the downlink user throughput distribution through online learning of non-linear interactions of features and lower bias learning models.  
In simulation, the proposed method outperforms industry standards in a realistic and scalable heterogeneous cellular environment.

\end{abstract}
\begin{IEEEkeywords} MIMO, DL CoMP, New Radio, NR, 5G, LTE-A, FDD, deep learning, heterogeneous networks, SON.\end{IEEEkeywords}

%
\IEEEpeerreviewmaketitle

\section{Introduction}
The aggregate demand for data traffic over \textit{fifth generation of wireless communications} (5G) cellular networks is expected to increase a thousand times compared to the previous generation \cite{6824752}.
Heterogeneous networks, in which small cells are deployed along with macro cell base stations, are one of the most important solutions to increase the network capacity.

Downlink \textit{coordinated multi-point} (DL CoMP) will play an important role in improving data rates and cellular capacity in 5G by using a centralized unit to coordinate the operation of multiple New Radio (NR) base station units  \cite{3gpp38804}.

DL CoMP (CoMP from now on) has various implementations. Our focus is on the \textit{joint transmission} scheme, where the \textit{user equipment} (UE) spatially multiplexed data streams are available at more than one point participating in the data transmission in a time-frequency resource.  These points (or base stations) form the CoMP \textit{cooperating set}.  This effectively forms a distributed \textit{multiple input multiple output} (MIMO) channel with decorrelated streams from each \textit{base station} (BS) in the CoMP cooperating set \cite{3gpp36819}.  {\color{black} These base stations communicate over low-latency backhaul.} A common approach in CoMP today is to use a static absolute triggering threshold.

\begin{figure}[!t]
\centering
\includegraphics[scale=0.55]{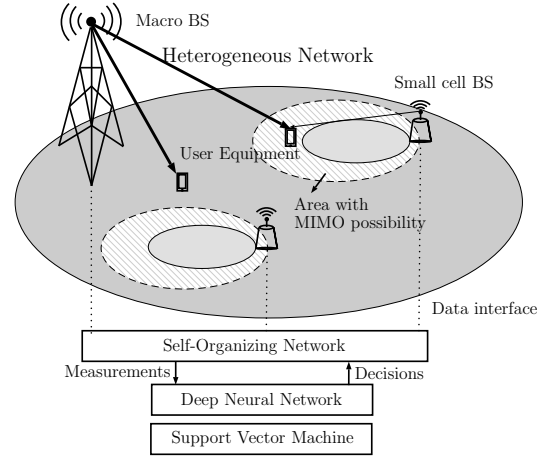}
\caption{Joint transmission in a coordinated multipoint New Radio heterogeneous network {\color{black} with interfaces to the self-organizing network.}}
\label{fig:dlcomp}
\end{figure}

In this letter, we further improve the CoMP joint transmission average user throughput performance from our previous work \cite{mysvm_paper}.  In our previous work, we used \textit{support vector machine} (SVM) binary classification for our CoMP trigger function in a \textit{frequency division duplex} (FDD) mode, which does not have channel reciprocity.  We propose an online deep learning algorithm which acquires physical layer measurement reports from the connected UEs within the channel coherence time in an FDD radio frame.  Our algorithm is compliant with the industry standard for CoMP for \textit{Long Term Evolution Advanced} (LTE-A) \cite{3gpp36819}. 
The proposed algorithm formulates a modified CoMP trigger function to enhance the downlink capacity.
The algorithm computation can take place in a centralized location as part of a \textit{self-organizing network} (SON) as shown in Fig.~\ref{fig:dlcomp}.  Our choice of \textit{deep neural networks} (DNNs) allows the creation of more learning features than shallow architectures such as SVM.  This is due to the combinatorial and non-linear nature of the hidden layers of a DNN.  Furthermore, DNNs perform particularly well when channels are complicated \cite{8052521,8400482}.  {\color{black} Also, SVMs tend to underperform when the classification problem is imbalanced \cite{svm-imbalance}.}

{\color{black} We choose a heterogeneous network due to the relatively shorter distances of small cells from the macro, making backhaul more suitable for CoMP \cite{3gpp36819}.  However, using macro BSs only may be possible with certain backhaul constraints~\cite{3gpp36819}.}

\vspace*{-1em}
\section{System Model} \label{sec:system_model}

\subsection{Network Environment}\label{sec:nw_environment}
{\color{black} Our heterogeneous network is based on \textit{orthogonal frequency division multiple} (OFDM) in the FDD mode of operation.  We uniformly scatter} small cells for densification of the macro coverage in an urban environment.  Non-stationary UEs with multiple antennas are randomly placed and uniformly distributed in the service area.  The network is comprised of a macro BS with one tier of surrounding macro BSs.  All macro BSs have three sectors with directional antennas.  We also add uniformly scattered small cells in the service area with omni-directional antennas.  The BSs are transmitters and the UEs are the receivers.  We use 5G NR as a multi-access wireless network in the sub-6 GHz range. 

\subsection{Signal Model}\label{sec:signal_model}
We write the received signal of an arbitrary UE $i$ as
\begin{equation}
\mathbf{r}_i = \mathbf{H}_i\mathbf{s}_i + \mathbf{v}_i, \qquad i = 1, \ldots, N_\text{UE}.
\label{eq:channel}
\end{equation}
Here, $\mathbf{r}_i  \in \mathbb{C}^{n_r}$, $\mathbf{H}_i\in \mathbb{C}^{n_r\times n_t}$ is a Rayleigh fading channel with independent and identically distributed circularly symmetric standard complex Gaussian entries, $\mathbf{s}_i \in \mathbb{R}^{n_t}$ is the transmitted signal, and $\mathbf{v}_i \in \mathbb{C}^{n_r}$ is the noise plus interference.  The latter two quantities are also assumed to be circularly symmetric Gaussian with zero mean and variance $\sigma_v^2\mathbf{I}$.  Also, $n_t$ and $n_r$ are the number of transmit and receive receive streams respectively such that the maximum number of streams $n_s^\textrm{max}\triangleq\min(n_r,n_t)$.

Since 5G NR is based on OFDM, we choose a \textit{zero-forcing} (ZF) combiner at the receiver (i.e., the UE).  This sets inter-cellular interference to zero thereby making the \textit{signal to noise ratio} (SNR) and \textit{signal to interference and noise ratio} (SINR) interchangeable. We write our ZF combiner for a given UE $i$ as $\mathbf{W}_{\text{ZF},i}\in\mathbb{C}^{n_t\times n_r}$ 
\begin{equation}
\mathbf{W}_{\text{ZF},i} = (\mathbf{H}_i^* \mathbf{H}_i)^{-1}\mathbf{H}_i^*
\end{equation}
Here, $(\cdot)^*$ is the Hermitian transpose operation.
The parameters of the radio environment are listed in Table~\ref{table:parameters}.

We compute the $i$-th UE SNR per receive stream $\gamma_j^{(i)}$:  
\begin{equation}
    \begin{aligned}
    \gamma_j^{(i)} = \frac{G_j^{(i)}P_{\text{BS},j}^{(i)}\Vert \mathbf{d}_i \Vert ^{-\alpha} }{n_t \sigma_v^2} / [\mathbf{H}_i^*\mathbf{H}_i]^{-1}_{j,j}, \qquad j = 1, \ldots, n_s
    \end{aligned}
    \label{eq:sinr}
\end{equation}
where $G_j^{(i)}$ is the corresponding large scale channel gain, $P_{\text{BS},j}^{(i)}$ is the transmit power of the base station in the CoMP cooperating set transmitting the $j$-th branch, $\mathbf{d}_i$ is the minimum distance of the $i$-th UE from the serving base station, $\alpha$ is the path loss exponent, all for the $i$-th UE.

 We also define the reference symbol received power for the $i$-th UE (measured on the first receive branch as)
\begin{equation}
    \begin{aligned}
    P_\text{RS}^{(i)} = G_\text{TX} P_{\text{BS}}^\textrm{max}\Vert \mathbf{d}_i \Vert ^{-\alpha}n_s \big / N_\text{SC}N_\text{PRB}
    \end{aligned}
    \label{eq:rsrp}
\end{equation}
where $G_\text{TX}$ is the base station antenna gain, $P_{\text{BS}}^\textrm{max}$ is the maximum base station transmit power, $N_\text{SC}$ is the number of subcarriers per \textit{physical resource block} (PRB) in the OFDM radio frame, and $N_\text{PRB}$ is the number of PRBs.

We define $\beta_{j,i}$, which is the respective codeword reception error, based on the \textit{block error rate} (BLER) for the $j$-th stream of the $i$-th UE.  BLER has a direct relationship with the decorrelation of streams and their chosen modulation and code scheme. We introduce two NR physical measurements which we use as learning features each of size $M$: $a$) the CSI \textit{reference symbol received power} (CSI-RSRP) which is \eqref{eq:rsrp} and $b$) a staircase transformation of the signal to interference and noise ratio of the data channel (CSI-SINR).  This resembles the \textit{channel quality indicator} (CQI) \cite{3gpp36213} from LTE-A and is the name we adopt here. We choose CQI and CSI-RSRP because they are two physical channel measurement quantities that are not correlated:  CSI-RSRP is the received power of the narrowband NR reference symbols on the first receive antenna only while CQI is a quantized indication of the received wideband SINR regardless of the receive antenna $j$ \cite{3gpp38215}.   The true CoMP-triggering function is unknown.

\vspace*{-.15in}
\subsection{Deep Learning} \label{sec:machinelearning}

\begin {table}[!t]
\setlength\doublerulesep{0.5pt}
\caption{Machine learning features for CoMP Improvement}
\vspace*{-0.1in}
\label{table:mlfeatures}
\centering
\begin{tabular}{c|lll} 
\hhline{====}
 & Parameter & Type & Description \\
 \hline
$\mathbf{x_1}$ & CSI-RSRP & Float & Narrowband received power measurement \\
$\mathbf{x_2}$ & CQI & Integer & Wideband received SINR linearly mapped \\
\hhline{====}
\end{tabular}
\end{table}

\begin{figure}[!t]
\centering
\begin{tikzpicture}[thick,scale=0.7, every node/.style={scale=0.7,font=\Large},   cnode/.style={draw=black,fill=#1,minimum width=3mm,circle},
]
 \node at (3,-4) {$\vdots$};
  \node at (6,-4) {$\vdots$};
    \foreach \x in {1,4}
    {   
    
      \pgfmathparse{\x== 3 ? "\vdots" : "\x"}
            \pgfmathparse{\x== 4? "2" : "\pgfmathresult"}
        \node[cnode=gray!20,label=180:$\mathbf{x_{\pgfmathresult}}$] (x-\x) at (0,{-\x-int(\x/5)-0.5}) {};
    }
    
    \foreach \x in {1,...,4}
    {
        \pgfmathparse{\x<4 ? \x : "w"}
        \node[cnode=gray] (x2-\x) at (3,{-\x-int(\x/4)}) {};
        \node[cnode=gray] (p-\x) at (6,{-\x-int(\x/4)}) {};
        
            }
    \node at (4.5,-0.75) {$\ldots$};
    \node at (4.5,-4.75) {$\ldots$};
    \node at (3,-0.5) {$\boldsymbol{\Theta}_1$};
    \node at (6,-0.5) {$\boldsymbol{\Theta}_d$};
   	
	\foreach \x in {3}
	{
	\pgfmathparse{\x== 3? "y" : "\pgfmathresult"}
        	\node[draw=black,circle,label=0:$\mathbf{\hat\pgfmathresult}$] (s-\x) at (9,{-\x-int(\x/3)+1 }) {};
	}
		    
    \foreach \x in {1,...,4}
    {   
           \draw (p-\x) -- (s-3);
        
    }

    \foreach \x in {1,...,4}
    {   \foreach \y in {1,...,4}
        {  \pgfmathparse{\x<4 ? \x : "H"}   
	        \draw (x2-\x) -- (p-\y) ; 
        }
    }
    \foreach \x in {1,4}
    {   \foreach \y in {1,...,4}
        {  \draw (x-\x) -- (x2-\y);
        }
    }
\end{tikzpicture}
\caption{Structure of the deep neural network used in the implementation of our modified algorithm.}
\label{fig:dnn}
\end{figure}
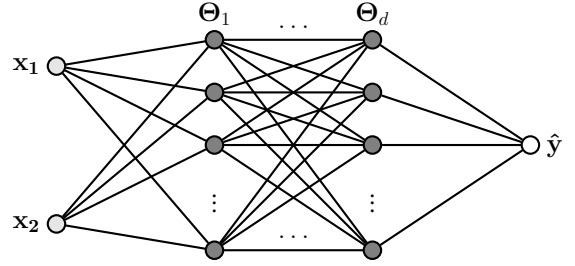

We use a DNN classifier with the \textit{rectified linear unit} (ReLU) activation function in the implementation of this algorithm as shown in Fig.~\ref{fig:dnn}.  The block matrix of the weights of its hidden layers is $\boldsymbol{\Theta} := [\boldsymbol{\Theta}_\ell]_{\ell = 1}^{d+1}, \boldsymbol{\Theta}_\ell\in\mathbb{R}^{w\times M}$.  We define the learning features in a matrix $\mathbf{X}$ based on the physical measurements in the previous subsection.  These features (listed in Table~\ref{table:mlfeatures}) are scaled such that their values lie in the closed interval $[0,1]$.  If the quantities were correlated, we would have seen an inflation in the training error variance rendering machine learning results useless.

To create the supervisory signal labels vector $\mathbf{y}\in\{0,1\}^M$, we use the aggregate BLER for the UE $i$, $\beta_i$, and write
\begin{equation}
    y_i := \mathbbm{1}_{\beta_{i} \le \beta_\text{target}}
    \label{eq:blertarget}
\end{equation}
where $\beta_\text{target}$ is the retransmission target. The vector $\mathbf{y}$ is likely to be imbalanced in the two classes as a result.  The aggregate BLER per user $i$, $\beta_i$, can be calculated from the transmission rank $n_s$ and the BLER per stream $j$, $\beta_{j,i}$, as
\begin{equation}
    \beta_i := 1-\prod_{j=1}^{n_s}(1-\beta_{j,i}).
    \label{eq:bler}
\end{equation}

The gathered data $\mathbf{X}$ and $\mathbf{y}$ is periodically split to a training and a test dataset.  We then train the model and tune the hyperparameters in Table~\ref{table:hyperparams}. We use grid search over the hyperparameters search space and $K$-fold cross-validation to prevent under- or over-fitting.

Since we train the DNN classifier with the training dataset, the anticipated generalization performance of the DNN classifier is represented by the misclassification error $\xi \triangleq  \sum_m \mathbbm{1}_{\hat y_m\neq \hat y_{\text{test}, m}} / M_\text{test}$, where $M_\text{test}$ is the test data size.
\vspace*{-1em}
\subsection{Problem Formulation}
A common approach to enable CoMP or disable it in the cooperating set is to use static absolute thresholds of quantities such as SINR.  However, these thresholds are subjective and are therefore unlikely to yield an optimal DL CoMP performance.  We formulate the problem of obtaining a dynamic CoMP threshold.  To do so, we collect radio measurements from all the UEs served by the cooperating set during the time duration of $T_\text{CoMP}$.  This duration cannot exceed the channel coherence time $T_\text{coherence}$ or the radio frame duration $T_\text{RF}$. {\color{black}  $T_\text{coherence} \approx c/v f_c$ in an OFDM transmission where $c$ is the speed of light, $v$ is the speed of the receiver, and $f_c$ is the center frequency of the OFDM carrier}. In other words, $T_\text{CoMP} \le \min(T_\text{coherence},T_\text{RF})$.  Given this, the matrix $\mathbf{X}$ has a number of rows $M$ upper bounded by $n_s^\textrm{max}N_\text{UE}gT_\text{CoMP}$, given the CSI reporting periodicity of $g$ reports per TTI as in \cite{3gpp38211}.  

The collected data is then used to train a deep learning classifier and if its performance is acceptable, it can override the static approach.  Otherwise, the static CoMP trigger is always the fallback.  The DNN classifier performance is measured through the decision threshold $\varepsilon$, which can also control misclassifications due to training outside the channel coherence time or poor model fitting in general.

The DNN classifier must be periodically invalidated (i.e., purged and retrained with new measurements) at a periodicity of $T_\text{CoMP}$.  Otherwise, the \textit{channel state information} (CSI) may have changed but have not been reflected onto the classifier.

We write the downlink channel capacity for an arbitrary stream $j$ as $C^{(j)} = \log_2(1 + \gamma_j)$ where $\gamma_j$ is \eqref{eq:sinr}.  Let $Z_i$ be equal to the BLER-penalized capacity for a given UE $i$, then
\begin{align}
 Z_i(n_s) :=
\begin{cases}
      \sum_j C^{(j)}(1-\beta_{i}), & \qquad \beta_{i} \le \beta_\text{target}\\
      0, & \qquad \text{otherwise}
\end{cases}
\label{eq:capstream}
\end{align}
where codewords are discarded if the receive error is above the retransmission target.  We employ a technique based on deep learning to estimate whether retransmission target will be met.  From \eqref{eq:blertarget} and \eqref{eq:capstream}, we find that $Z_i := C_iy_i$ where $C_i$ is the sum of the capacity across all streams $n_s$. We therefore build a deep learning classifier where $y_i := [Z_i/C_i]$.  This enables us to reformulate the problem as a machine learning problem minimizing the binary cross-entropy loss function $L(\cdot,\cdot)$:
\begin{equation}
\underset{\boldsymbol{\Theta}}{\text{minimize:}} \,\, L(\mathbf{y},\mathbf{\hat y}) := -\sum_k y_k\log \hat y_k + (1-y_k)\log(1-\hat y_k)
\label{eq:problem}
\end{equation}
where $\mathbf{\hat y}$ is estimated from the DNN classifier.    The value of $\mathbf{\hat y}$ instructs the CoMP cooperating set to form or teardown a dynamic MIMO channel through changing $n_s$.  This is done per user for all users $i$ during a given \textit{transmission time interval} (TTI).  Deep learning transforms \eqref{eq:problem} to higher dimensions through combinatorial and non-linear nature of the hidden layers.  {\color{black} Using the DNN in Fig.~\ref{fig:dnn}, we write our \textit{surrogate function} $\mathbf{\hat y}$ in terms of the two inputs, the trainable weights $\boldsymbol{\Theta}$, and non-linear activation functions $\sigma_\ell(\cdot)$ as
\begin{equation}
    \mathbf{\hat y} = \sigma_{d+1}(\boldsymbol{\Theta}_{d+1}\sigma_d(\ldots\sigma_1(\boldsymbol{\Theta}_1\mathbf{\tilde X})))
    \label{eq:surrogate}
\end{equation}
where $d+1$ is the output layer, $\mathbf{\tilde X}$ is the normalized matrix $\mathbf{X}$, and the non-linear activation functions are applied element-wise on vectors.  It can be inferred from the surrogate function formula that the number of learning features generated from these inputs is in $\mathcal{O}(w^d)$.}  This surrogate function is a CoMP-triggering function used for the next $T_\text{CoMP} - 1$ TTIs.

\vspace*{-1em}
\subsection{Performance Measurement}
The peak (95th percentile), average, and edge (5th percentile) of the user downlink throughput empirical distribution \cite{VLS-2016} are used as a measurement of performance. 

\vspace*{-1em}
\section{Algorithms}\label{sec:algorithm}
\begin{enumerate}[leftmargin=*]
    \item \textit{Static DL CoMP Algorithm}: The decision to enable or disable CoMP in the cooperating set for users is static and based on an absolute threshold of the DL SINR reported by the distribution of users.
    \item \textit{Dynamic DL CoMP Algorithm}: The dynamic algorithm to trigger CoMP comes from \cite{mysvm_paper}. The asymptotic time complexity of SVM training is in $\mathcal{O}(M^3)$ where $M$ is the number of rows in the matrix $\mathbf{X}$ as computed in Section~\ref{sec:system_model}.
    \item \textit{Deep Learning DL CoMP Algorithm}: The improved proposed dynamic algorithm to trigger CoMP is shown in Algorithm~\ref{alg:algocomp}. The lower bound time complexity of training a DNN with $d$ hidden layers and $w$ neurons per hidden layer is in $\mathcal{O}(Mw^d)$ but can also run in parallel \cite{scikit-learn}.  Otherwise, with an equal hyperparameter search space size and cross-validation fold size, DNN run-time complexity outperforms SVM if $d\log w<2\log M$.
\end{enumerate}

\begin{algorithm}[!t]
\small
 \DontPrintSemicolon
 \KwIn{Decision threshold $\varepsilon$, measurements collection period $T_\text{CoMP}$, current triggering DL SINR. Table~\ref{table:simparameters} has example values.}
\KwOut{Triggering decision for DL CoMP for all $N_\text{UE}$ UEs in $T_\mathrm{sim}$ TTIs.}
\SetKwBlock{Begin}{procedure}{end procedure}

  \For {$T := 1$ \KwTo $T_\mathrm{sim}$} {
     Acquire the learning features $\mathbf{X}$ in Table~\ref{table:mlfeatures} from all UE measurements during time $t = T,\ldots,(T + T_\text{CoMP} - 1)$. \;
     Compute the classification label $\mathbf{y}$. \;
    \If {$T\!\!\mod T$\textsubscript{\rm CoMP}$ = 0$} {
         Split the measurement data $[\mathbf{X}\,\vert\, \mathbf{y}]$ to training and test data. \;
         Scale the features in $\mathbf{X}$ to interval $[0,1]$.\;
         Train the DNN model using the training data and use grid search on $K$-fold cross-validation to tune the hyperparameters (in Table~\ref{table:hyperparams}) and compute $\mathbf{\hat{y}}$. \;
         Compute the misclassification error $\xi$. \;
          
        \eIf {$\xi \le \varepsilon$}  { 
          
            Decision is to override setting and enable DL CoMP in next TTI if $\textrm{mean}(\mathbf{\hat{y}}) \ge 0.5$ else disable DL CoMP in next TTI. \;
        }  
        {            
              Fallback to operator-entered DL CoMP SINR trigger (static algorithm).\;
         }
        Invalidate the DNN model. \;
        Purge collected measurement data.\;
    }
}
\caption{\small Deep Learning DL CoMP Algorithm}\label{alg:algocomp}
\end{algorithm}

\vspace*{-1em}
\section{Simulation Results} \label{sec:simulation}

\begin{table*}[!t]
\setlength\doublerulesep{0.5pt}
\centering
\caption{Classifier hyperparameters}
\vspace*{-.1in}
\label{table:hyperparams}
\begin{threeparttable}
\vspace*{-.1in}
\begin{tabular}{ll | ll}
\hhline{====}
DNN Hyperparameter & Search range & SVM Hyperparameter & Search range \\
\hline
DNN depth $d$ & \{1,3,5\} & Kernel & \{\texttt{gaussian, polynomial}\tnote{\textasteriskcentered}\;\} \\ 
DNN width $w$ & \{{\color{black} 3,10}\} & Kernel scale and Box constraint & \cite{mycode_dnn} \\
\hhline{====}
\end{tabular}
\begin{tablenotes}\footnotesize
\item[\textasteriskcentered]{Degrees $p\in \{1,2, 3, 4\}$.}
\end{tablenotes}
\end{threeparttable}
\end{table*}

\begin{table}[!t]
\setlength\doublerulesep{0.5pt}
\caption{Radio Environment Parameters}
\vspace*{-.1in}
\label{table:parameters}
\centering
\begin{tabular}{ lr}
\hhline{==}
Parameter & Value \\
 \hline
NR Bandwidth $B$ & 10 MHz \\
Downlink center frequency $f_c$ & 2100 MHz \\
Macro BS maximum power & 46 dBm \\
Small cell BS maximum power & 37 dBm  \\
Maximum number of streams $n_s^\textrm{max}$  & 2 \\
Number of PRBs $N_\text{PRB}$ & 50 \\
\hhline{==}
\end{tabular}
\end{table}

We use a MATLAB-based simulator to implement our algorithm \cite{mycode_dnn,VLS-2016}.   The UEs are at an average speed $v = 5\, \text{km/h}$.  We use a $K=5$ $K$-fold cross-validation with a training-test data split of 70-30.  The retransmission target $\beta_\text{target}$ is set to 10\%, the number of subcarriers per PRB $N_\text{SC} = 12$, and the radio frame duration $T_\text{RF} = 10$ ms. The important simulation parameters are in Table~\ref{table:simparameters}. 

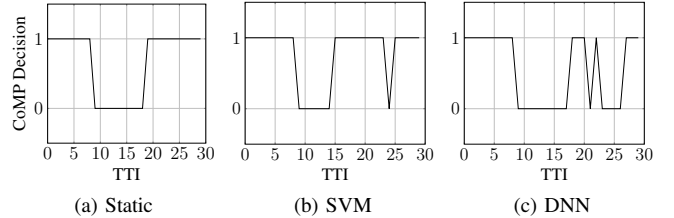
\begin{figure}[!t]
\centering
\subfloat[Static]{\resizebox{1.1in}{!}{
%
%
\begin{tikzpicture}
\tikzstyle{every node}=[font=\Large,scale=3.5]
\begin{axis}[%
width=6.028in,
height=5.304in,
at={(1.011in,0)},
scale only axis,
xmin=0,
xmax=30,
xlabel={TTI},
xmajorgrids,
ymin=-0.5,
ymax=1.5,
xtick={0,5,...,30},
ytick={0,1},
ylabel={CoMP Decision},
ymajorgrids,
axis background/.style={fill=white},
]
\addplot [color=black,solid,forget plot]
  table[row sep=crcr]{%
0	1\\
1	1\\
2	1\\
3	1\\
4	1\\
5	1\\
6	1\\
7	1\\
8	1\\
9	0\\
10	0\\
11	0\\
12	0\\
13	0\\
14	0\\
15	0\\
16	0\\
17	0\\
18	0\\
19	1\\
20	1\\
21	1\\
22	1\\
23	1\\
24	1\\
25	1\\
26	1\\
27	1\\
28	1\\
29	1\\
};
\end{axis}
\end{tikzpicture}%
}}
\hfil
\subfloat[SVM]{\resizebox{1.1in}{!}{
%
%
\begin{tikzpicture}
\tikzstyle{every node}=[font=\Large,scale=3]
\begin{axis}[%
width=6.028in,
height=4.754in,
at={(1.011in,0.642in)},
scale only axis,
xmin=0,
xmax=30,
xlabel={TTI},
xmajorgrids,
ymin=-0.5,
ymax=1.5,
xtick={0,5,...,30},
ytick={0,1},
ymajorgrids,
axis background/.style={fill=white},
]
\addplot [color=black,solid]
  table[row sep=crcr]{%
0	1\\
1	1\\
2	1\\
3	1\\
4	1\\
5	1\\
6	1\\
7	1\\
8	1\\
9	0\\
10	0\\
11	0\\
12	0\\
13	0\\
14	0\\
15	1\\
16	1\\
17	1\\
18	1\\
19	1\\
20	1\\
21	1\\
22	1\\
23	1\\
24	0\\
25	1\\
26	1\\
27	1\\
28	1\\
29	1\\
};
\end{axis}
\end{tikzpicture}%
}}
\hfil
\subfloat[DNN]{\resizebox{1.1in}{!}{
%
%
\begin{tikzpicture}
\tikzstyle{every node}=[font=\Large,scale=3]
\begin{axis}[%
width=6.028in,
height=4.754in,
at={(1.011in,0.642in)},
scale only axis,
xmin=0,
xmax=30,
xlabel={TTI},
xmajorgrids,
ymin=-0.5,
ymax=1.5,
xtick={0,5,...,30},
ytick={0,1},
ymajorgrids,
axis background/.style={fill=white},
]
\addplot [color=black,solid]
  table[row sep=crcr]{%
0	1\\
1	1\\
2	1\\
3	1\\
4	1\\
5	1\\
6	1\\
7	1\\
8	1\\
9	0\\
10	0\\
11	0\\
12	0\\
13	0\\
14	0\\
15	0\\
16	0\\
17	0\\
18	1\\
19	1\\
20	1\\
21	0\\
22	1\\
23	0\\
24	0\\
25	0\\
26	0\\
27	1\\
28	1\\
29	1\\
};
\end{axis}
\end{tikzpicture}%
}}
\caption{Downlink coordinated multipoint being enabled \mbox{(state = 1)} and disabled (state = 0) for the static (left), the SVM-based (middle), and the DNN proposed algorithm (right).} 
\label{fig:sinrcomp6}
\end{figure}

In Fig. \ref{fig:sinrcomp6}, the static algorithm makes decisions to enable or disable CoMP in the cooperating set when the improved dynamic algorithms do the opposite.  The performance improvement for the peak and average cases is shown in Table~\ref{table:improvements}.  This improvement is due to the learning of an improved surrogate CoMP triggering function.  The reason for DNN outperforming the SVM-based CoMP algorithm is {\color{black} two-fold}: First, the depth of the DNN allows the creation of more features.  {\color{black} The number of features in DNN is $\mathcal{O}(w^d)$, compared to the most feature-generating polynomial SVM kernel of degree $p$ with the number of features being $\mathcal{O}(p)$. Second, SVM tends to suffer bias towards the majority class when the training supervisory signal labels are imbalanced (i.e., $\#(y_i = 0)\!=\!$ 1,522 out of 9,180) \cite{svm-imbalance}. Due to the decisions made by the DNN computed surrogate function \eqref{eq:surrogate}, the CoMP cooperating set prevents UEs from receiving codewords with higher BLER penalty \eqref{eq:bler}.  Furthermore, the cooperating set deactivates CoMP at times where SVM decision is biased towards enabling CoMP (or vice versa). The optimization of triggering CoMP with a reduced BLER and a smaller number of streams $n_s$ brings about the observed downlink throughput gain.}
Neither the average CQI nor the average RSRP was impacted as shown in Table~\ref{table:linklevel}.

\begin{table}[!t]
\setlength\doublerulesep{0.5pt}
\caption{Simulation Parameters}
\vspace*{-.2in}
\label{table:simparameters}
\begin{center}
\begin{threeparttable}
\begin{tabular}{ lr } 
\hhline{==}
Parameter & Value \\
 \hline
Static DL CoMP SINR trigger  & -3.5 dB \\
Total number of connected UEs in the cluster $N_\text{UE}$ & 180 \\
Number of small cells & 17 \\
Number of macro BSs & 21 \\
Measurements collection period $T_\text{CoMP}$ & 3 TTIs \\
Simulation time $T_\text{sim}$ & 30 TTIs \\
Misclassification error threshold $\varepsilon$ & 15\% \\
 \hhline{==}
\end{tabular}
\end{threeparttable}
\end{center}
\end{table}

\begin {table}[!t]
\setlength\doublerulesep{0.5pt}
\caption{Downlink Throughput Improvement over Static CoMP}
\vspace*{-.1in}
\label{table:improvements}
\centering
\begin{tabular}{ p{0.1\textwidth}ccc}
\hhline{====}
Percentile & Static [Mbps] & SVM [Mbps] & DNN  [Mbps] \\
 \hline
Peak (95\%) & 2.84 & 3.08 (0.9\%) & \textbf{3.29 (15.8\%)} \\
Average & 1.02 & 1.10 (7.8\%) & \textbf{1.16 (13.7\%)} \\
Edge (5\%) & 0.07 & 0.07 (0.0\%) & \textbf{0.08 (14.3\%)} \\
\hhline{====}
\end{tabular}
\end{table}

\begin{table}[!t]
\setlength\doublerulesep{0.5pt}
\caption{Downlink Link-level Average Measures}
\vspace*{-.1in}
\label{table:linklevel}
\centering
\begin{tabular}{ p{0.1\textwidth}cccc} 
\hhline{=====}
Algorithm & BLER $\beta_i$ & Streams $n_s$ & CQI & CSI-RSRP [dBm] \\
 \hline
SVM CoMP & 7.15\% & 1.59 & 3 & -58.17\\
DNN CoMP & \textbf{6.73}\% & \textbf{1.55} & 3 & -58.17\\
\hhline{=====}
\end{tabular}
\end{table}

\section{Conclusions}\label{sec:conclusion}
In this letter, we motivated the use of a surrogate trigger function for CoMP.  We obtained this function through applying online deep learning to physical layer measurements in a realistic NR FDD heterogeneous network.   Our standards-compliant method using DNN enhanced downlink rates compared to SVM with virtually no impact on the reported CQI or the narrowband received power.  This improvement is due to the increased number of features and the lower bias in the DNN classification compared to SVM.
\vspace*{-.1in}






%
%
%
\bibliography{references}  
\vspace*{-2in}
\bibliographystyle{IEEEtran}

%

%
%




\end{document}